\documentclass{pasa}

\usepackage{amsmath,amssymb}
\usepackage{hyperref} 
\usepackage{aas-macros}
\usepackage{graphicx}
\usepackage{xspace}
\usepackage[table,xcdraw]{xcolor}
\usepackage{mdframed}
\usepackage{subcaption}
\usepackage{amsmath}
\usepackage{balance}
\usepackage{multirow}
\hypersetup{colorlinks=true,citecolor=blue,linkcolor=blue,urlcolor=magenta}

\title{Gravitational pulsars: correlations between the electromagnetic and the continuous gravitational wave signal}

\author[]{Marco Antonelli$^1$\thanks{\url{antonelli@lpccaen.in2p3.fr}}
\href{https://orcid.org/0000-0002-5470-4308}{\includegraphics[scale=.5]{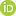}},
Avishek Basu$^2$\thanks{\url{avishek.basu@manchester.ac.uk}}
\href{https://orcid.org/0000-0002-4142-7831}{\includegraphics[scale=.5]{ORCID-iD_icon-16x16.png}},
Brynmor Haskell$^{3,4,5}$\thanks{\url{bhaskell@camk.edu.pl}}
\href{https://orcid.org/0000-0002-8255-3519}{\includegraphics[scale=.5]{ORCID-iD_icon-16x16.png}}
\affil{$^1$ CNRS/in2p3, Laboratoire de Physique Corpusculaire (LPC Caen), 14050 Caen, France}
\affil{$^2$ Jodrell Bank Centre for Astrophysics, School of Physics and Astronomy, University of Manchester, M13 9PL, Manchester, UK}
\affil{$^3$ Dipartimento di Fisica, Universit\`{a} di Milano, Via Celoria 16, 20133 Milano, Italy}
\affil{$^4$ INFN, Sezione di Milano, Via Celoria 16, 20133 Milano, Italy}
\affil{$^5$ Nicolaus Copernicus Astronomical Center, Polish Academy of Sciences, ul. Bartycka 18, Warszawa, Poland}
}

\jid{PASA}
\doi{10.1017/pas.\the\year.xxx}
\jyear{\the\year}

\begin{document}


\begin{frontmatter}
\maketitle

\begin{abstract}
Neutron stars emitting continuous gravitational waves may be regarded as gravitational pulsars, in the sense that it could be possible to track the evolution of their rotational period with long-baseline observations of next-generation gravitational wave interferometers.
Assuming that the pulsar's electromagnetic signal is tracked and allows us to monitor the pulsar's spin evolution, we provide a physical interpretation of the possible observed correlation between this timing solution and its gravitational counterpart, if the system is also detected in gravitational waves. In particular, we show that next-generation detectors, such as the Einstein Telescope, could have the sensitivity to discern different models for the coupling between the superfluid and normal components of the neutron star and constrain the origin of timing noise (whether due to magnetospheric or internal processes). Observational confirmation of one of the proposed scenarios would therefore provide valuable information on the physics of gravitational wave emission from pulsars.
\end{abstract}

\end{frontmatter}


\section{Introduction}

In the dense and cold interiors of mature neutron stars (NSs), nuclear matter is expected to become superfluid~\citep{Chamel2017JApA,Sedrakian2019EPJA}. This can have astrophysical implications, as superfluids can sustain permanent bulk currents (thus introducing additional macroscopic degrees of freedom to the hydrodynamic description~\citealt{haskell_sedrakian2018ASSL,thermo_2020CQG}), quantized vortices and distinct sound modes~\citep{graber2017IJMPD,nils2021Univ}.

The timing of pulsars (rapidly rotating magnetized NSs) in radio, X- and $\gamma$-rays has already provided insights into the interior dynamics of a superfluid component~\citep{alessandro_review_1996,amp_review_2023}. 
In particular, pulsar glitches -- sudden spin-up events observed in over a hundred pulsars -- are believed to result from the abrupt exchange of angular momentum between the interior superfluid and the `normal', observable, component of the star~\citep{Zhou2022Univ,Antonopoulou2022review,amp_review_2023}. Additionally, the longer-term variations to the pulsar's spin, known as intrinsic timing noise (TN), may be associated with fluctuations in the internal torques between the components~\citep{alpar_noise_1986, jones_noise_1990MNRAS, MelatosLink2014, Lower2021mnras}, in the external spin-down torques~\citep{Lyne+2010,Brook2018aho,meyers2021a,Shaw+2022,Basu:2024qwt}, or possibly both~\citep{antonelli_timing2023}.

The rotation rate of the internal component remains, however, impossible to access with electromagnetic (EM) data. 
The situation may be different in the future, as Gravitational Wave (GW) detections are starting to become commonplace. 
While signals from compact binary inspirals involving NSs have already been detected by the LIGO-Virgo-KAGRA (LVK) network (see \citealt{GWTC3} for a recent catalogue of detected GW events), for the current analysis we are interested in a different class of signals, that has not yet been detected, but is being actively searched for, Continuous Gravitational Waves (CWs). These are long-lived, quasi-monochromatic signals, thought to originate mostly (although not only) from rotating NSs \citep{Piccinni2022Galax,Riles2023LRR}, essentially the GW equivalent of a pulsar. 

There are several mechanisms that can allow a rotating neutron star to emit a continuous GW signal. They can primarily be classed into `static' (in the rotating frame) perturbations, or `mountains', that are swept around by rotation and thus emit at multiples of the rotation frequency \citep{GittinsReview}, and modes of oscillation that have an intrinsic frequency, which itself can depend on the rotation frequency (as is the case for the $r$-mode, \citealt{haskell2015IJMPE}). 

The above description generally assumes that all components of the star rotate together. However, the presence of superfluid phases in the interior forces us to consider two or more components rotating differentially. It is therefore natural to ask whether the GW signal tracks the observable rotation frequency of the `normal' component or that of the superfluid, which may account for a significant fraction of the star's moment of inertia.
For instance, single-harmonic narrow-band searches are commonly conducted near twice the pulsar's rotation frequency~\citep{Piccinni2022Galax}, thus implicitly accounting for a potential discrepancy in rotation rates between the magnetosphere region responsible for pulsations and the bulk region responsible for the CW emission.
 
Going beyond rigid rotation is complex, and a number of works have been devoted to studying this problem for different scenarios \citep{jones2004_timing, HasAnd09, Glampedakis09, Hogg13, ashton_timing_2015, haskell2024APh}, with the answer possibly depending on still poorly understood phenomena that are believed to affect the rotation of pulsars, such as the dynamics of superfluid vortices \citep{antonelli2020MNRAS,levin2023ApJ,liu2025ApJ} and internal turbulence \citep{anders2007MNRAS,MelatosLink2014}.

It would therefore be of great interest to observationally track both the CW and EM emission from a pulsar, as this would provide indirect information about the spin evolution of the internal superfluid regions that sustain the strongest persistent currents, and how closely they follow the magnetosphere and the parts strongly coupled to it (such as the normal component or superfluid regions where strong permanent currents cannot develop). Such observations could help investigate GW emission mechanisms and the coupling between the superfluid and normal components of the star (see, e.g.,~\citealt{antonelli2020MNRAS,levin2023ApJ}), as well as offer insights into the physics of pulsar glitches and timing noise.
 
To date, no CW has yet been detected, but searches in O3 data have started to constrain astrophysically significant parameter space for several sources \citep{HasBejger23}, and future detectors, such as the Einstein Telescope (ET) or Cosmic Explorer, will likely be sensitive to a large number of such sources in our galaxy~\citep{Branchesi_2023}.

With the sensitivity of future GW detectors in mind, it is thus important to address the relevance of pulsar timing noise for CW searches, and the relationship between the NS's EM and CW signals. \citet{jones2004_timing} carried out a first analysis of this problem, investigating the relationship between the phase residual from the GW emission ($\delta \Phi_{GW}$) and the EM radiation  ($ \delta \Phi_{EM}$). 
Three possibilities were identified, which naturally arise in different idealized scenarios, giving rise to three different benchmark values for the `slope' parameter $\alpha = \delta \Phi_{GW}/ \delta \Phi_{EM}$, which is a measure of the correlation between the two phase residuals.

First, the two signals may be strongly coupled, resulting in identical phase evolution (the timing noise in the detected CW signal matches the EM one, $\alpha\approx1$). 
This would be the case if the normal fluid and the superfluid are coupled on very short timescales compared to the observation time or the detector's sensitivity to resolve variations occurring over such brief periods.

The second possibility is that the two signals may be loosely coupled and show similar levels of timing noise but their phases do not match, $\alpha\approx0$. This case may also be realised when the EM signal has timing noise, but the CW signal does not have any phase wandering.

Finally, if the mechanism giving rise to the phase wandering in the GW and EM signals is such that each signal is in phase with the rotation rate of an internal component, then the total angular momentum conservation will impose the constraint $x_{GW}\delta \dot{\Phi}_{GW}(t)+x_{EM}\delta \dot{\Phi}_{EM}(t)=0$, where $x_{GW}$ and $x_{EM}$ are the moments of inertia fractions of the internal components tracked by the EM (the magnetosphere) and the GW (e.g., the deformed crust or the internal superfluid) signals. In this case, it is natural to expect anti-correlation with a typical slope parameter of~$\alpha=-x_{GW}/x_{EM}$.

In this paper, we analyse the spin-wandering and take an agnostic approach regarding the exact GW  emission mechanisms. We consider the case of a hypothetical future detection in which we have both EM observations that allow us to time the normal component, and CW observations that track an internal component, not locked to the normal one. 
Given this scenario, in Sec.~\ref{section2} we show that information can be extracted from the comparison between the power spectral density (PSD) of the EM and CW signals or — without going to the frequency domain — from the slope $\alpha$. Sec.~\ref{sec3} is devoted to the numerical validation of our analytical calculation of $\alpha$ and to verifying that our stochastic toy model can reproduce the peak-to-peak amplitudes of TN for both canonical and millisecond pulsars. Finally, in Sec.~\ref{gwdetection}, we examine whether there are promising sources that could be detected as `gravitational pulsars'. We draw our conclusions in Sec.~\ref{sezioncione_discussiolone}.

\section{Correlation and spectral properties}
\label{section2}
 
We consider a two-component neutron star with angular velocities $\mathbf{\Omega}_t = \left(\Omega_p(t), \Omega_n(t)\right)$, where one component (labelled $p$) represents the EM observable `normal' component (protons, electrons, and all constituents strongly coupled to the magnetosphere), and the other (labelled $n$) corresponds to the neutron superfluid, as is standard in glitch models~\citep[see][for a review]{Haskell2015IJMPD,amp_review_2023}.

Following \citet{meyers2021a}, we model the spin-down of the NS in terms of the stochastic equation
\begin{equation}
\label{langevinTOT}
\dot{\mathbf{\Omega}}_t = B \, \mathbf{\Omega}_t + 
\mathbf{A} + M \dot{\mathbf{W}}_t  \, , 
\end{equation}
where $\dot{\mathbf{W}}_t$ is a vector consisting of two independent standard white noise processes, 
\begin{equation}
\label{Wnoise}
\langle \dot{\mathbf{W}}_t \rangle  = 0 \, ,
\qquad \quad    
\langle \dot{W}^i_t \dot{W}^j_s\rangle  = \delta_{ij} \delta(t-s) \, ,
\end{equation}
and
\begin{equation}
\label{2_comp_mat}
B = \begin{pmatrix}
    -\dfrac{x_n}{\tau} & \dfrac{x_n}{\tau} \\[1em]
    \dfrac{x_p}{\tau} & -\dfrac{x_p}{\tau}
     \end{pmatrix} ,
\quad 
M = \begin{pmatrix}
    \dfrac{\sigma_\infty}{x_p}  &  -\dfrac{\sigma_\mathcal{T}}{x_p} \\[1em]
    0                  &   \dfrac{\sigma_\mathcal{T}}{x_n}
     \end{pmatrix} ,
\end{equation}
with
\begin{equation}
\mathbf{A}= \begin{pmatrix}
-\dfrac{\dot{\Omega}}{x_p}\\[1em]
    0
     \end{pmatrix} .
\end{equation}
In the above expressions $x_n$ is the moment of inertia fraction of the superfluid responsible for the internal fluctuating torque, $x_p=1-x_n$ is the fractional moment of inertia of the rest of the star, $\tau$ is the relaxation timescale due to internal friction, while $\sigma_\mathcal{T}$ sets the intensity of the fluctuations in the internal mutual friction torque, and $\sigma_\mathcal{\infty}$ sets the strength of fluctuations in the external braking torque~\citep{antonelli_timing2023}. Finally, $\dot{\Omega}>0$ is the absolute value of the average spin-down rate, measured over a long period.

To study long-term fluctuations due to timing noise, we are interested in the timing residuals, i.e. the deviations from a deterministic spin-down model. We therefore define the angular velocity residuals $\delta\mathbf{\Omega}_t=\left(\delta{\Omega}_p(t) , \delta{\Omega}_n(t)\right)$ as
\begin{equation}
\label{ang_vel_res}
     \delta\mathbf{\Omega}_t = \mathbf{\Omega}_t  -  \langle \mathbf{\Omega}_t \rangle \, ,
\end{equation}
where the expected value $\langle \mathbf{\Omega}_t \rangle$ coincides with the steady-state deterministic drift of $\mathbf{\Omega}_t$ due to $\mathbf{A}_t$, namely
\begin{equation}
     \langle \mathbf{\Omega}_t \rangle =\begin{pmatrix}
   \Omega-\dot{\Omega} \, t\\
   \Omega+\tau\dot{\Omega}/x_p-\dot{\Omega}\,t
     \end{pmatrix} .
\end{equation}
The evolution of the angular velocity residuals is then given by the stochastic equation:
\begin{equation}
\label{langevin2}
\delta\dot{\mathbf{\Omega}}_t =  B \, \delta\mathbf{\Omega}_t +M\dot{\mathbf{W}}_t \, .
\end{equation}
The above equation provides a simple framework to derive the expected features of timing noise, like its power spectral density (PSD) matrix
\begin{equation}
\label{rice00}
     P_{ij}(\omega) = \left[ \left( i \omega\,\mathbb{I} - B \right)^{-1} M M^\top 
     \left( -i \omega \,\mathbb{I}- B^\top \right)^{-1} \right]_{ij}\, , 
\end{equation}
where $i,j=n,p$ (the PSD matrix is Hermitian since the residuals are real). 
The existence of internal superfluid layers leaves its mark on the PSD of the two components~\citep{antonelli_timing2023}, obtained by setting $i=j=p$ and $i=j=n$ in the above equation:
\begin{eqnarray}
&& P_{p}(\omega) = \dfrac{x_p^2 \sigma_\infty^2 + \omega^2 \left(\sigma_\mathcal{T}^2+\sigma_\infty^2\right) \tau^2}{ \omega^2  x_p^2 \left(1+ \omega^2 \tau^2\right)}
\propto \frac{\omega^2+\mu_p^2}{\omega^2+\xi^2} \,  \omega^{ -\beta} ,
\nonumber \\
&& P_{n}(\omega) = \dfrac{x_n^2 \sigma_\infty^2+\omega^2 \sigma_\mathcal{T}^2 \tau^2}{ \omega^2  x_n^2 \left(1+ \omega^2 \tau^2\right)}\propto \frac{\omega^2+\mu_n^2}{\omega^2+\xi^2} \, \omega^{ -\beta} ,
\nonumber \\
&&\xi=\frac{1}{\tau}\,  , 
\qquad \quad
\mu_n = \frac{x_n  \sigma_{\infty} }{\tau \,  \sigma_\mathcal{T} } \, ,
\nonumber \\
&&\mu_p =  \frac{x_p   \sigma_{\infty} }{\tau \left(\sigma_\mathcal{T}^2+\sigma_{\infty}^2\right)^{1/2} } \, .
\label{psds}
\end{eqnarray}
%
In the present specific case where white noise is injected, we have that $\beta=2$. It is possible, however, to directly inject red noise and obtain a higher spectral index $\beta>2$ for the asymptotic decay of $P_n$ and $P_p$. For example, if instead of injecting $\dot{\mathbf{W}}_t$ (white noise) we were to inject  $\mathbf{W}_t$ (Wiener noise), we would obtain exactly the two PSDs in \eqref{psds} but with $\beta=4$, recovering a more red behaviour that is similar to the one obtained with the alternative stochastic model for timing noise proposed by~\citet{vargas2023MNRAS}. An explicit example is carried out in App.~\ref{app1}.

Whatever the injected noise model, the main message is that the presence of an internal superfluid component leaves its imprint in the PSD $P_{p}$ of the residuals $\delta\Omega_p$ as a whiter\footnote{
    At low ($\omega \ll \mu_p$) and high ($\omega \gg \xi$) frequencies the PSD for $\delta\Omega_p$ behaves as $\omega^{-\beta}$. In between the two corner frequencies, the PSD behaves as $\omega^{2-\beta}$, which is a flat spectrum typical of white noise if $\beta=2$.}
region $\mu_p<\omega<\xi$ defined by the two corner frequencies $\mu_p$ and~$\xi$. Note that the corner frequencies are always ordered as~$0 < \mu_p <\xi$, see equation~\eqref{psds}.

Similarly, if $\delta\Omega_n(t)$ were observable in the GW channel, we could cross-validate the presence of the corner frequency $\xi$ in $P_n$ and, possibly, of~$\mu_n$. In this case, however, there is no unique possible ordering of the corner frequencies: if $0<\mu_n<\xi$ then the intermediate frequency region tends to be whiter ($P_n\sim \omega^{2-\beta}$), if $0<\xi<\mu_n$ then the intermediate region is redder~($P_n\sim \omega^{-2-\beta}$).

Possible extraction of $\delta\Omega_n(t)$ from CW data would still be valuable even without transitioning to the frequency domain, which would require sufficiently high resolution to distinguish the corner frequencies $\xi$ and $\mu_n$. An alternative to studying $P_n$ could be to extract information from the correlation between the CW and EM signals, as measured by the parameter~$\alpha$~\citep{jones2004_timing}.
By imposing the initial condition $\delta\mathbf{\Omega}_{t=0}=0$ at a certain arbitrary time $t=0$, we solve the stochastic system \eqref{langevin2} and compute the equal-time correlation matrix 
\begin{equation}
c^{ij}_{t}=\langle \delta{\Omega}_t^i  \delta{\Omega}_t^j \rangle  \, , 
\end{equation}
where the brackets indicate the average over noise realizations. Explicit calculation gives
\begin{equation}
    c^{ij}_t = c_0^{ij} + c_1^{ij} e^{-t/\tau} + c_2^{ij} e^{-2t/\tau} \approx c_0^{ij}
    \quad(\text{for}\,\,\,t\gg\tau)
\end{equation}
where $c_1^{ij}$ and $c_2^{ij}$ are constant coefficients and
\begin{equation}
\begin{split}
& c_0^{pp}= t \, \sigma_\infty^2 -\frac{\tau}{2 x_p^2} \left( 
\sigma_\mathcal{T}^2 + \sigma_\infty^2 (1+2 x_p +3 x_p^2)\right) \, ,
\\
& c_0^{nn}=   \frac{\tau}{2 x_n^2} \left(
\sigma_\mathcal{T}^2 -3 \sigma_\infty^2  x_n^2 \right) \, ,
\\
& c_0^{np}= t \, \sigma_\infty^2 -\frac{\tau}{ x_n x_p} \left( 
\sigma_\mathcal{T}^2 + \sigma_\infty^2 x_n(3 x_p-1)\right) \, .
\end{split}
\end{equation}
This allows us to find the long-term average slope $\alpha$ of the trajectory $\delta{\mathbf{\Omega}}_t$  in the plane of the possible residual values $(\delta{\Omega}^n_t, \delta{\Omega}^p_t)$ as
\begin{equation}
    \alpha \, = \,  c^{np} \, / c^{pp} 
    \approx \,
    c_0^{np}/ c_0^{pp} 
    \quad(\text{for}\,\,\,t\gg\tau)
    \label{slopealpha}
\end{equation}
where the approximation is valid for long times. For $t \gg \tau$ (so that the system loses memory of the arbitrary initial condition $\delta\mathbf{\Omega}_{t=0}=0$), we find two interesting limits: $\alpha=1$ if there are no fluctuations in the internal torque (i.e., $\sigma_\mathcal{T}=0$), and  
$\alpha=-x_p/x_n$ when there are no fluctuations in the external torque~(i.e., $\sigma_\mathcal{\infty}=0$). 

\section{Synthetic timing noise}
\label{sec3}

In order to validate the above analysis and get a sense of the numbers involved, we simulate the stochastic dynamics in~\eqref{langevin2}. 
In the following, we will adopt the common parametrisation of the relaxation timescale $\tau$ in terms of the dimensionless friction parameter $\mathcal{B}$,
\begin{equation}
    \tau \, = \, x_p/(2\, \Omega \, \mathcal{B}) \, ,
\end{equation}
see \citep{montoli2020A&A} for of how to interpret this body-averaged parameter starting from models that account for local stratification and gradients. 
Moreover, rather than using the parameters $\sigma_\mathcal{T}$ and $\sigma_{\infty}$ to set the intensity of the internal and external torque fluctuations, we will use two dimensionless parameters $\alpha_\mathcal{T}$ and $\alpha_\infty$. For a given pulsar of angular velocity $\Omega$ and absolute value of the spin down rate $\dot{\Omega}$, they are defined via 
\begin{equation}
\sigma_\mathcal{T}^2 \,=\, \dfrac{ \alpha_\mathcal{T}^2 \, x_1^2  \,  
\dot{\Omega}^2 }{2 \, \mathcal{B} \,  \Omega}  \, ,
\qquad \qquad \quad
\sigma_\infty^2 \, = \, \alpha_\infty^2 \, \Omega \, \dot{\Omega} \, .   
\end{equation}
In this way, the dimensionless parameters $0\leq \alpha_\mathcal{T}<1$ and $0\leq \alpha_\infty<1$ set the relative strength of fluctuations with respect to the deterministic part of the corresponding torque~\citep{antonelli_timing2023}.

In view of the analysis outlined in Sec.~\ref{gwdetection}, the synthetic timing noise signal extracted from simulations corresponds to the rotational parameters $\Omega$ and $\dot{\Omega}$ of the canonical pulsar J2043+2740 and the millisecond pulsar J1024-0719.
In particular, we aim to reproduce the peak-to-peak timing noise amplitude\footnote{
    The peak-to-peak amplitude of TN over a time interval $T$ is an important feature to consider when assessing the potential detection of CWs during an observation of duration $T$, which will be considered in~Sec.~\ref{gwdetection}.
    } 
observed in J2043+2740 and J1024$-$0719 by integrating the stochastic system in equation~\eqref{langevinTOT} for different choices of physical parameters. This procedure yields the angular velocity residuals $\delta\boldsymbol{\Omega}_t$, or equivalently, the frequency residuals. The corresponding phase residuals, $\left(\delta \Phi_p(t), \delta \Phi_n(t)\right)$, and time-of-arrival (TOA) residuals - hereafter referred to simply as timing residuals - $\left(\delta a_p(t), \delta a_n(t)\right)$, are then computed using equations (76) and (85) of~\cite{antonelli_timing2023}. 
Within this scheme, the timing and phase residuals are proportional to each other and carry the same information. Nonetheless, we use both representations, as EM observations of timing noise often employ some measure of the typical amplitude of the timing residuals $\delta a_p(t)$ to quantify the TN strength, while searches for CW with interferometers generally rely on the phase wandering $\delta\Phi_n(t)$ of the signal~\citep{jones2004_timing,Piccinni2022Galax}.

\subsection{PSR J2043+2740} 


PSR J2043+2740 is the shortest period object in the sample of 17 canonical pulsars of \citet{Shaw+2022}, completing one rotation every $96\,$ms~(i.e., $\Omega\approx 65.4\,$rad/s, corresponding to $\nu_{GW}\approx20.8\,$Hz assuming GW emission from a non-precessing triaxial star). 
This pulsar exhibits a strongly correlated variation in both emission and the spin-down rate $\dot \Omega$ (with a change in $\dot \Omega\approx 8.4\times 10^{-13}$rad/s$^2$ of about $6\%$), occurring approximately once every 10 years \citep{Shaw+2022}.  
This $\sim6\%$ change in $\dot\Omega$ does not visibly affect the timing residuals we compare with, and is included in the behaviour reported by \citet{Shaw+2022}, which we use as a reference. This change is interpreted as evidence of magnetospheric variability, suggesting an external origin for at least part of the TN, which, within our framework, is expected to be more consistent with simulations driven by a fluctuating external torque.

To test whether our model can reproduce the peak-to-peak TN amplitude observed in this pulsar, we simulate \eqref{langevin2} in two limiting cases: one where the source of noise is almost entirely internal, and one where it is almost entirely external. 

\begin{figure*}
    \centering
    \begin{minipage}[t]{0.48\textwidth}
        \centering
        \includegraphics[width=\textwidth]{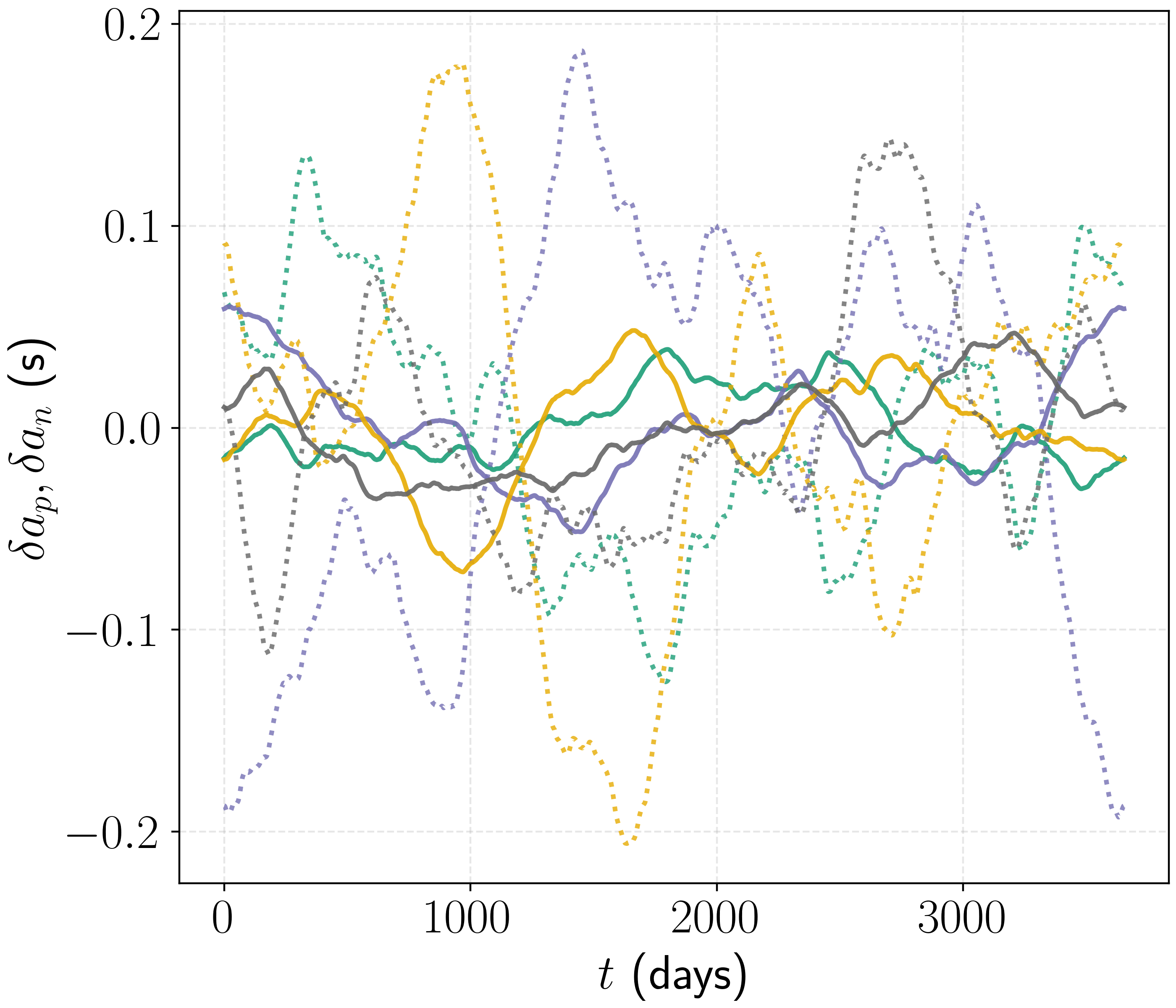}
    \end{minipage}%
    \hfill
    \begin{minipage}[t]{0.48\textwidth}
        \centering
        \includegraphics[width=\textwidth]{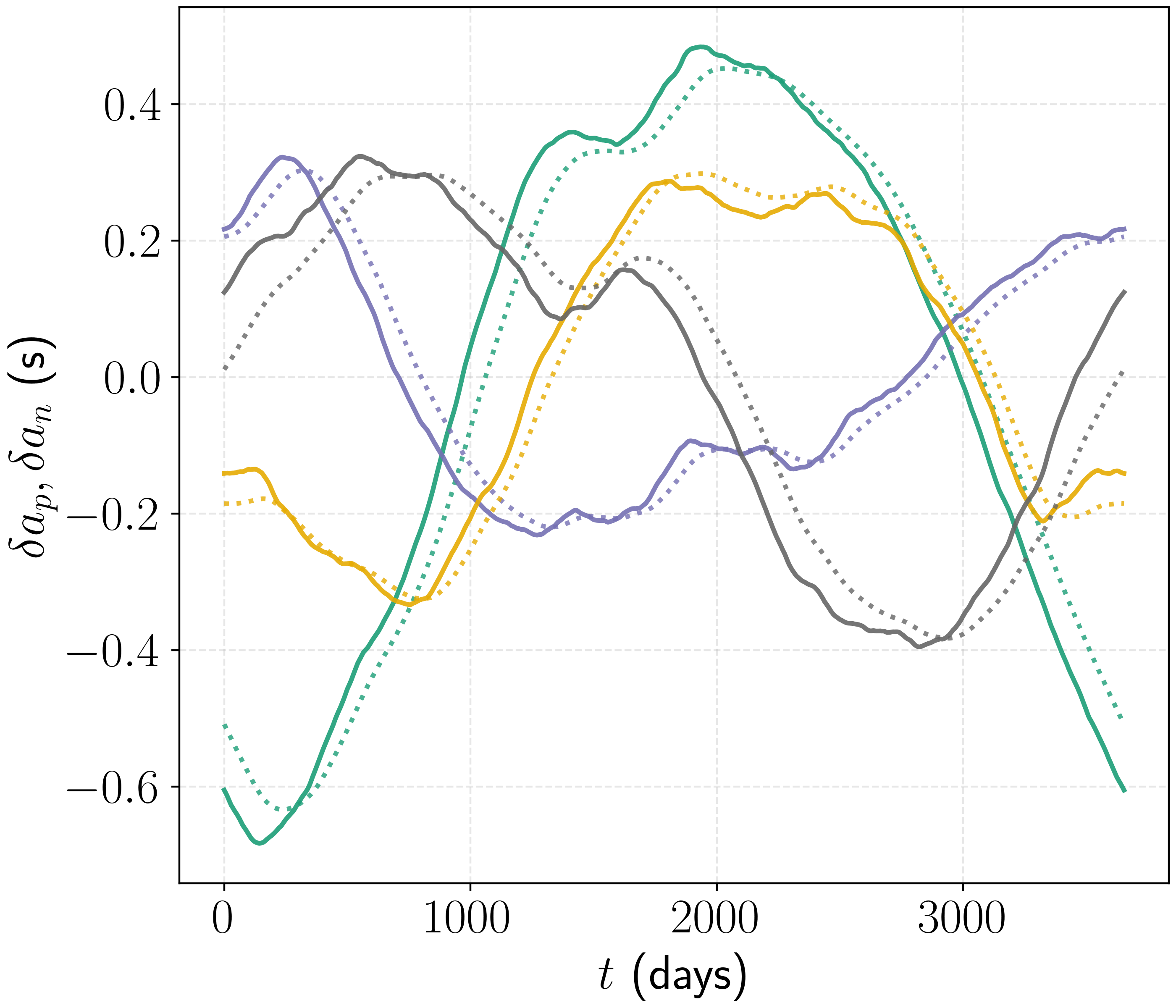}
    \end{minipage}
    \caption{
    Simulated timing residuals $\delta a_p(t)$ and $\delta a_n(t)$ for the canonical pulsar J2043+2740 (solid and dotted lines, respectively), shown for four noise realizations. 
    \emph{Left panel:} Internal noise scenario with parameters $x_p=0.8$, $\mathcal{B}=10^{-9}$, $\alpha_\mathcal{T}=10^{-1}$, $\alpha_\infty=10^{-6}$, yielding a relaxation time $\tau\approx 70\,$days. 
    Despite the large $\alpha_\mathcal{T}$, the simulated peak-to-peak amplitude underestimates the observed $\sim2\,$s variation over $\sim10$~yr reported in \citet{Shaw+2022}. 
    \emph{Right panel:} External noise scenario with $\alpha_\mathcal{T}=10^{-13}$ and $\alpha_\infty=10^{-5}$, keeping all other parameters the same. This configuration reproduces the observed TN peak-to-peak amplitude.
    }
    \label{fig:toa_2043_both}
\end{figure*}

According to \citet{Shaw+2022}, the peak-to-peak timing noise amplitude reaches approximately $2200\,$ms over $\sim10$ years. As shown in Fig.~\ref{fig:toa_2043_both}, the internal noise scenario (left panel), with $\alpha_\mathcal{T} = 0.1$ and $\alpha_\infty = 10^{-6}$, fails to reproduce the observed TN amplitude despite the relatively high level of internal torque fluctuations. In contrast, the external noise scenario (right panel), with $\alpha_\mathcal{T} = 10^{-13}$ and $\alpha_\infty = 10^{-5}$, successfully matches the observed variation in timing residuals. 
Hence, in our toy model, even small stochastic variations in the external torque can lead to large-amplitude TN features when the rotational parameters of PSR J2043+2740 are used, in line with the interpretation that the timing noise is primarily magnetospheric in origin for this pulsar.

\begin{figure*}
    \centering
    \begin{minipage}[t]{0.48\textwidth}
        \centering
        \includegraphics[width=\textwidth]{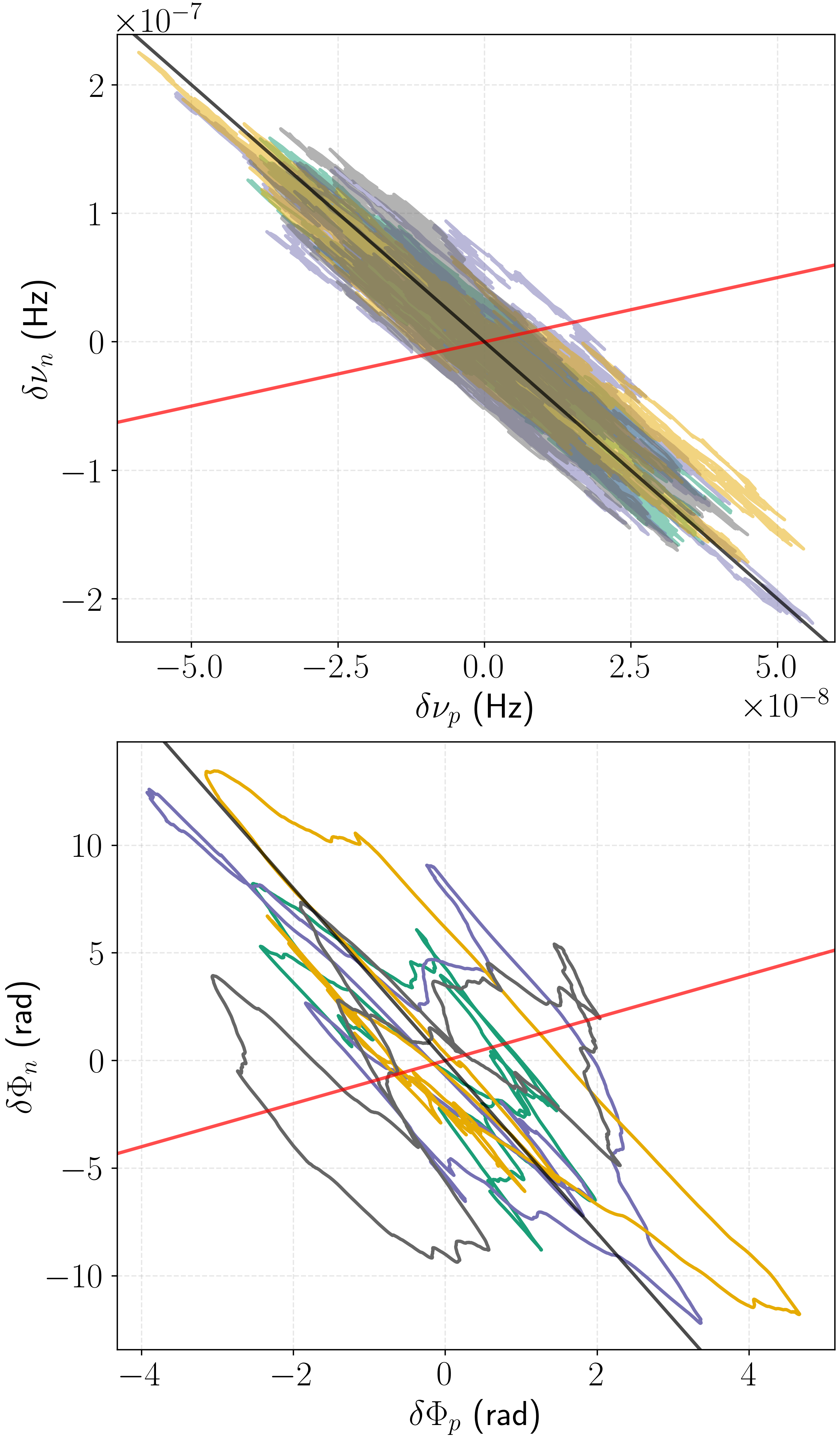}
    \end{minipage}%
    \hfill
    \begin{minipage}[t]{0.48\textwidth}
        \centering
        \includegraphics[width=\textwidth]{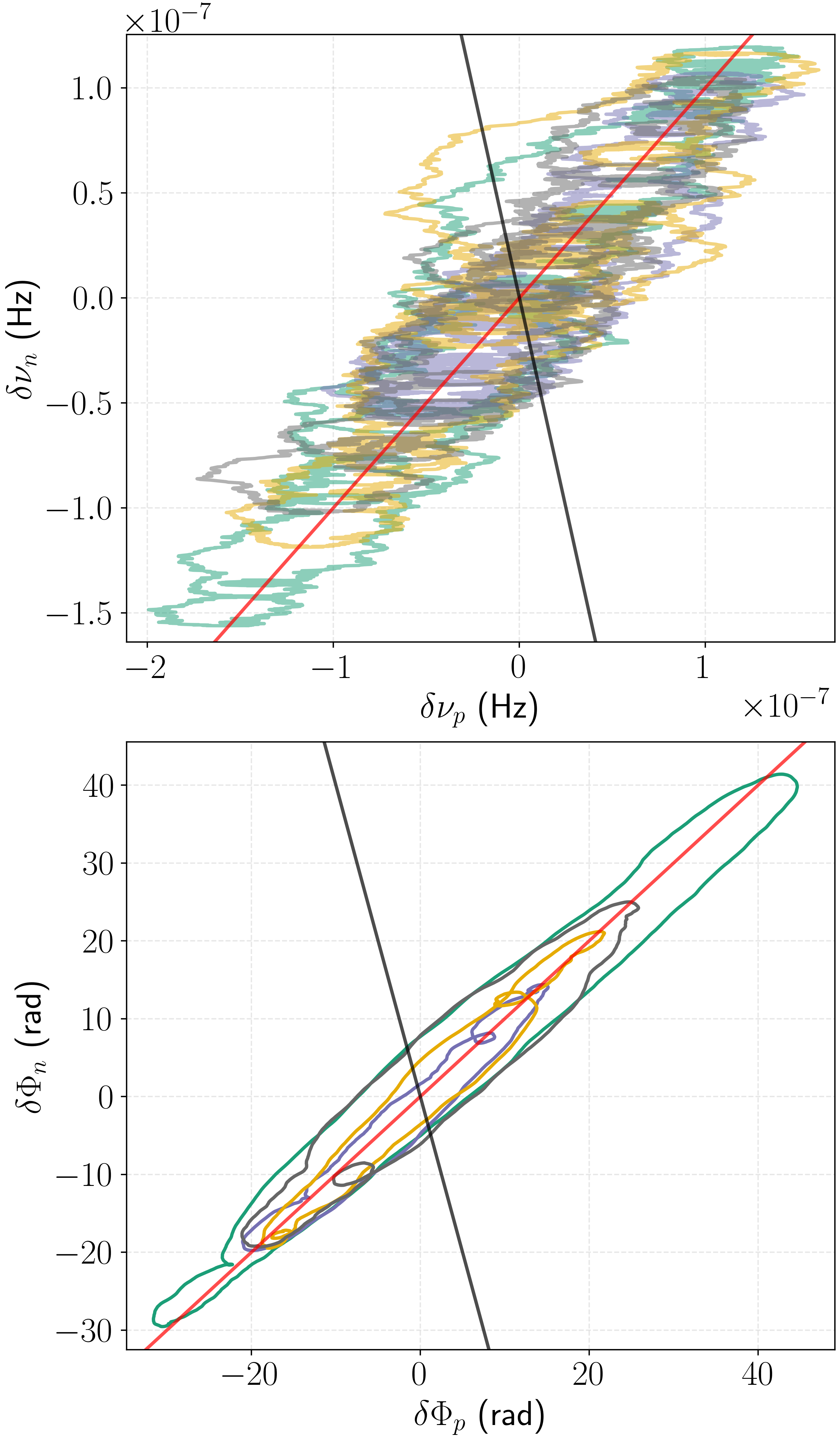}
    \end{minipage}
    \caption{
    Four examples of typical trajectories in the $(\delta \nu_p,\delta \nu_n)$ and $(\delta \Phi_p,\delta \Phi_n)$ planes for the non-recycled pulsar J2043+2740, corresponding to the same four noise realizations shown in Fig.~\ref{fig:toa_2043_both}. 
    \emph{Left panel:} Internal noise scenario. Although the noise originates mainly from internal torque, the presence of small external torque fluctuations causes the trajectories to deviate from the black line of slope $\alpha = -x_p/x_n = -4$ (for comparison, the red line has slope $\alpha = 1$).
    \emph{Right panel:} External noise scenario. In this case, a small internal torque noise component is still present, causing deviations from the red line with slope $\alpha = 1$ (the black line again marks $\alpha = -x_p/x_n = -4$).
    }
    \label{fig:corr_2043_both}
\end{figure*}

To assess the recoverability of the correlation slope $\alpha \approx \delta \Phi_n / \delta \Phi_p$ from synthetic data, we examine the trajectory of the residuals $(\delta\Omega_p(t), \delta\Omega_n(t))$, $(\delta\nu_p, \delta\nu_n)$, and $(\delta\Phi_p, \delta\Phi_n)$ for both internal and external noise scenarios. 
These are shown in Fig.~\ref{fig:corr_2043_both}, which corresponds to the same four noise realizations used in Fig.~\ref{fig:toa_2043_both}. In the internal noise case (left panel), the residual trajectories tend to cluster around a line with slope $\alpha = -x_p/x_n = -4$, as expected from the conservation of angular momentum in the absence of significant external torque noise. However, even in this limiting case, even small external fluctuations or the simple presence of a long relaxation timescale $\tau$ cause noticeable deviations from the ideal correlation line.

In the external noise case (right panel), the trajectories follow a line close to $\alpha = 1$, as expected from~\eqref{slopealpha}. Yet, the presence of a small internal noise contribution and the finite relaxation timescale $\tau$ still produces deviations from the ideal linear behavior. These deviations underscore an important point: even when the dominant noise source is well understood, accurately recovering the correlation slope $\alpha$ from data requires sufficiently long observational baselines to average out the stochastic spread. The relatively wide trajectory envelopes in both scenarios highlight the challenge of inferring the internal dynamics from phase residuals in any single realization.

Taken together, Figs.~\ref{fig:toa_2043_both} and \ref{fig:corr_2043_both} demonstrate that although idealized models provide good guidance, the interplay of internal and external noise — as well as finite coupling times — produces non-negligible scatter in realistic datasets. Careful statistical treatment will be required to extract meaningful information from future GW-resolved timing noise.

\subsection{PSR J1024-0719} 


Regarding millisecond pulsars, \citet{Perera+2019} analysed the timing noise of PSR J0437-4715, PSR J1024-0719 and PSR J1939+2134.
The one with the strongest timing noise is J1939+2134 (B1937+21). However, its timing residuals draw an almost perfect sinusoid of period 30 yr, likely due to a planetary companion or precession~\citep{Vivekanand2020ApJ}. For this reason, we focus instead on PSR J1024$-$0719~\citep{Kaplan2016ApJ}, which has the second-largest peak-to-peak amplitude in the TOA residuals among the MSPs studied by~\citet{Perera+2019}.

This pulsar has a rotational period of $P \approx 5.2$ ms (i.e., $\Omega \approx 1200$ rad/s), with long-term observations spanning over 18 years. The timing residuals exhibit a peak-to-peak variation of $\sim10^{-4}$ s, with an RMS of the TOA of about $7.3\,\mu$s~\citep{Perera+2019}.

\begin{figure*}
    \centering
    \begin{minipage}[t]{0.48\textwidth}
        \centering
        \includegraphics[width=\textwidth]{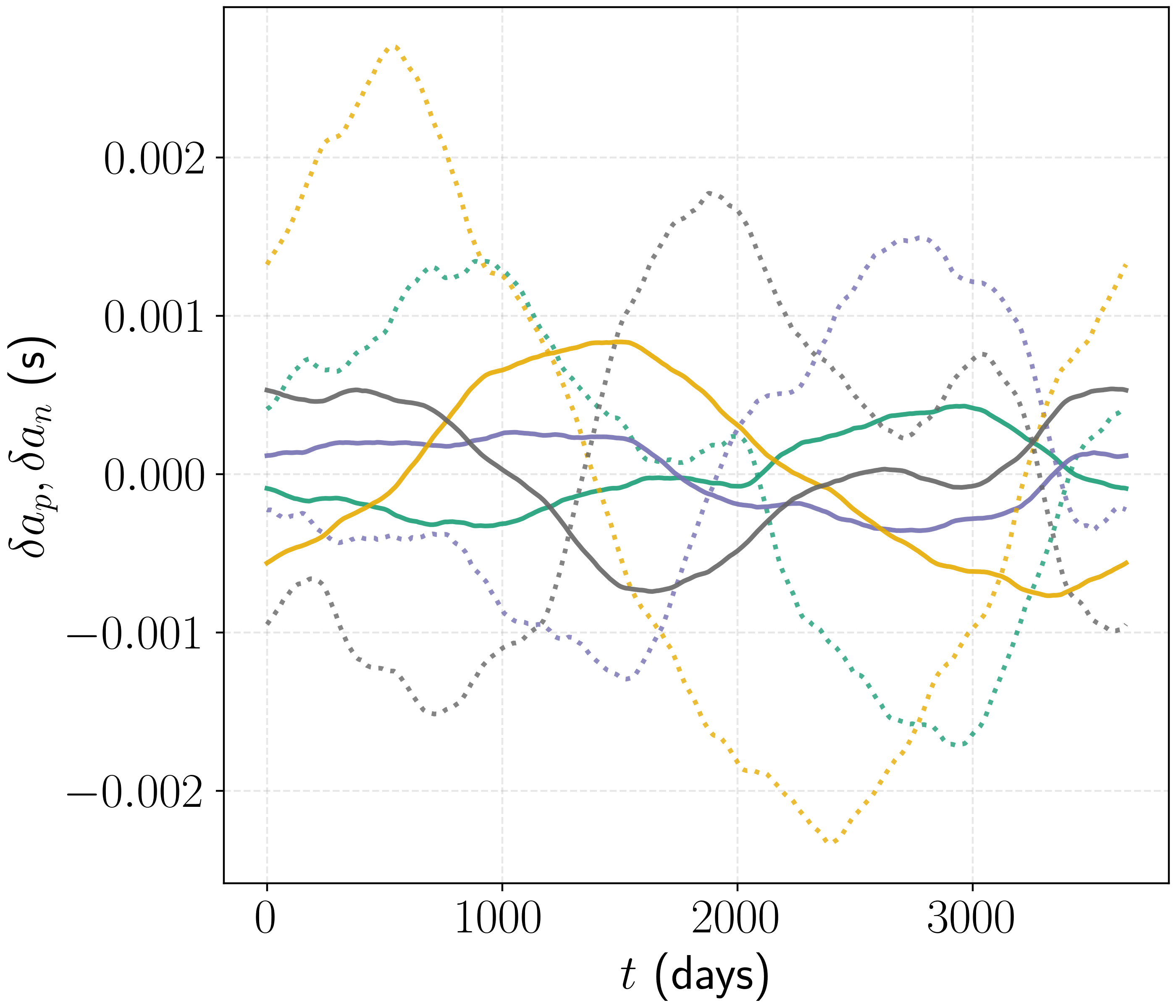}
    \end{minipage}%
    \hfill
    \begin{minipage}[t]{0.48\textwidth}
        \centering
        \includegraphics[width=\textwidth]{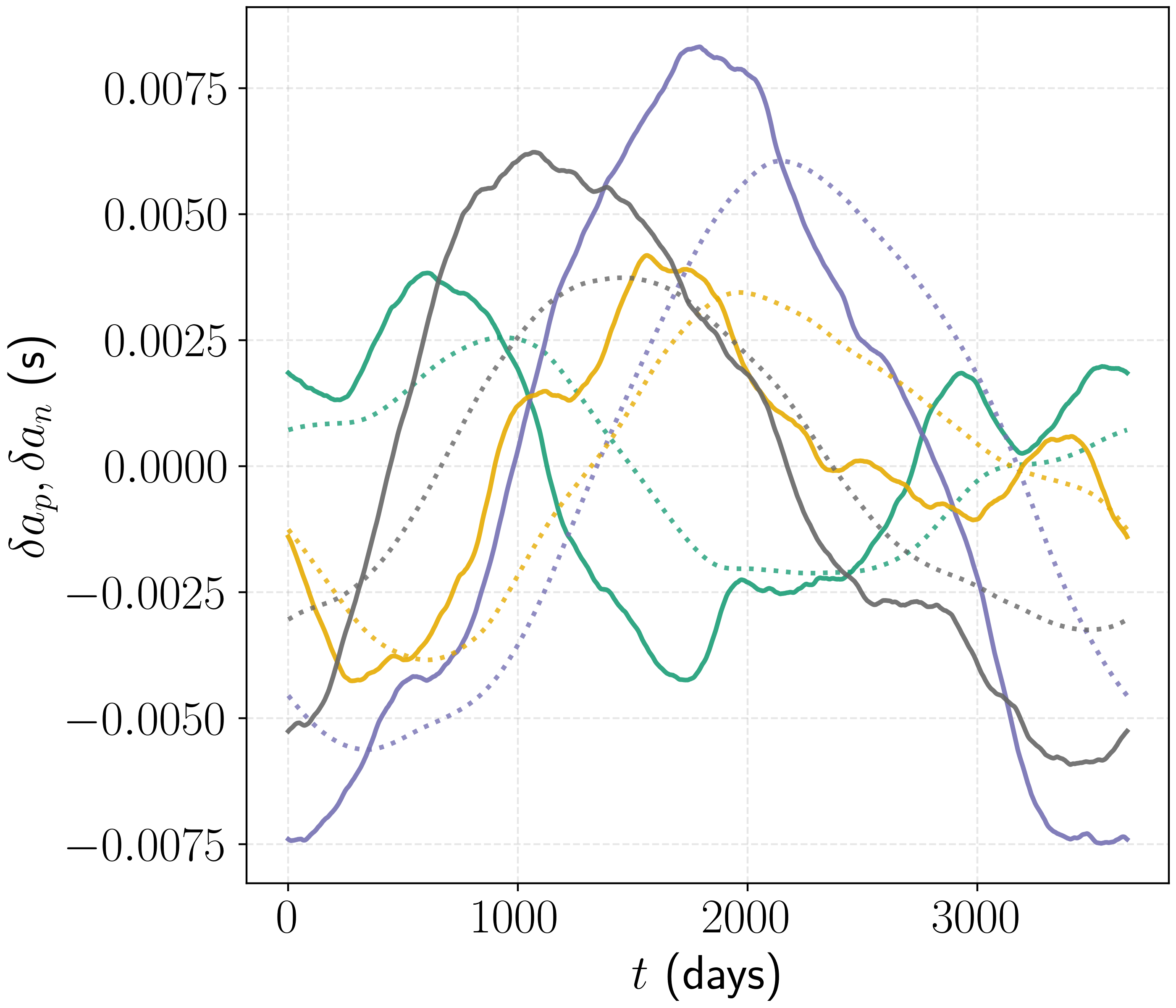}
    \end{minipage}
    \caption{
    Simulated timing residuals $\delta a_p(t)$ and $\delta a_n(t)$ for the millisecond pulsar J1024$-$0719 (solid and dotted lines, respectively), shown for four noise realizations. 
    \emph{Left panel:} Mostly internal noise scenario with parameters $x_p = 0.8$, $\mathcal{B} = 10^{-12}$, $\alpha_\mathcal{T} = 10^{-1}$, and $\alpha_\infty = 10^{-8}$, resulting in a relaxation time $\tau \approx 3800$~days. This configuration tends to overestimate the observed peak-to-peak timing residual amplitude of $\sim 10^{-4}$~s over $\sim 10$~yr, as reported in \citet{Perera+2019}.
    \emph{Right panel:} Mostly external noise scenario with $\mathcal{B} = 10^{-11}$, $\alpha_\mathcal{T} = 10^{-8}$, and $\alpha_\infty = 10^{-5}$ (all other parameters unchanged), which reproduces the observed TN amplitude. This supports an external (e.g., magnetospheric) origin for the residuals in this MSP.
    }
    \label{fig:toa_1024_both}
\end{figure*}

As for the canonical pulsar J2043+2740, we explore whether our two-component model can account for the observed timing noise amplitude in two contrasting regimes: one where the fluctuations are predominantly internal and one where they are mostly external. Fig.~\ref{fig:toa_1024_both} shows the simulated TOA residuals $\delta a_p(t)$ and $\delta a_n(t)$ for both scenarios. In the left panel, the internal noise case, we adopt $\alpha_\mathcal{T} = 10^{-1}$ and $\alpha_\infty = 10^{-8}$, yielding a long relaxation time of approximately $3800$ days (these parameter values are chosen as simple powers of ten for illustrative purposes and are not fine-tuned to reproduce the observed amplitude in detail). This long timescale is necessary to maximise the effect of the fluctuations in the internal torque, as it can be seen in the scaling of $P_p$ in \eqref{psds}. This configuration leads to timing noise amplitudes that systematically overshoot the observed variation over a 10-year span (a more realistic peak-to-peak amplitude can be achieved by tuning the $\mathcal{B}$ parameter). We deliberately retain this choice of a very long $\tau$ to test what happens when the relaxation timescale is comparable to the observational baseline (recall that \eqref{slopealpha} is valid in the opposite limit).

On the other hand, the external noise scenario (right panel), with $\alpha_\mathcal{T} = 10^{-8}$ and $\alpha_\infty = 10^{-5}$, matches more closely the observed peak-to-peak amplitude of $\sim 10^{-4}$ s: as in the case of the canonical pulsar, small but persistent stochastic variations in the external torque can drive long-term timing noise consistent with observations. Given the generally low levels of internal superfluid activity expected in MSPs (they tend not to exhibit glitches, and the few detected are of very small amplitude), this result seems to favour a magnetospheric (external) origin of the noise in PSR J1024$-$0719 as well.

\begin{figure*}
    \centering
    \begin{minipage}[t]{0.45\textwidth}
        \centering
        \includegraphics[width=\textwidth]{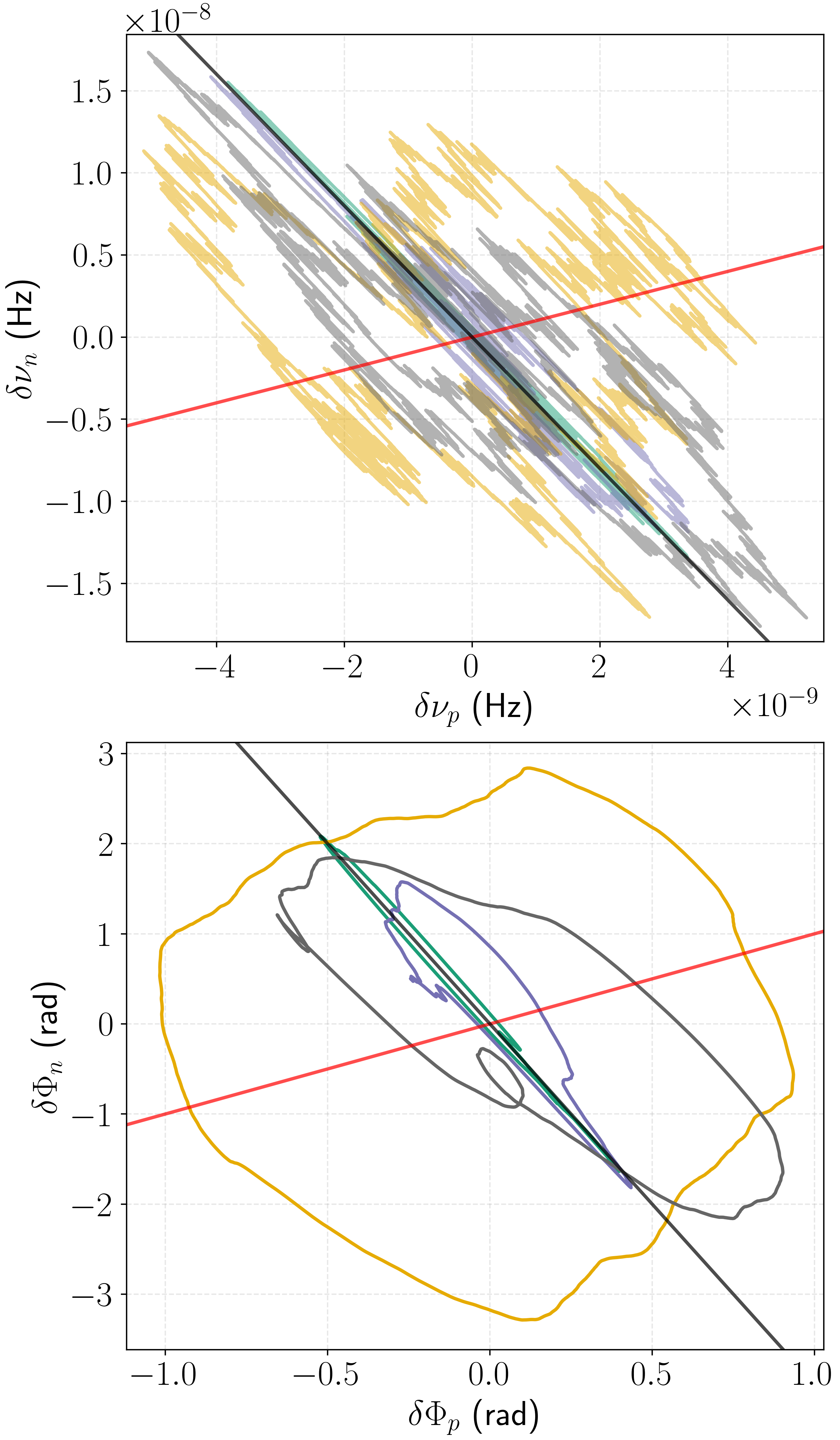}
    \end{minipage}%
    \hfill
    \begin{minipage}[t]{0.45\textwidth}
        \centering
        \includegraphics[width=\textwidth]{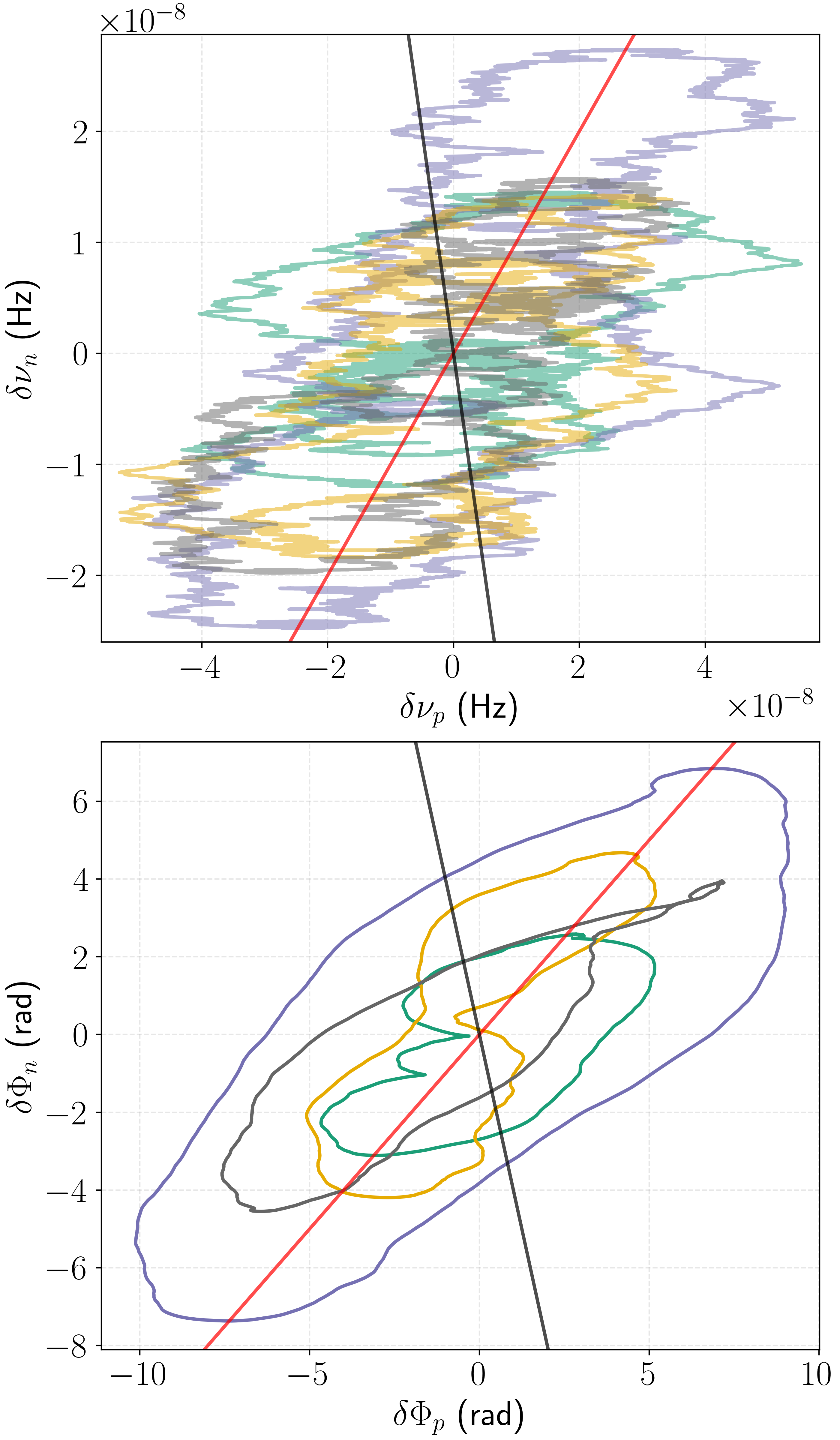}
    \end{minipage}
    \caption{
    Trajectories in the $(\delta \nu_p,\delta \nu_n)$ and $(\delta \Phi_p,\delta \Phi_n)$ planes for the millisecond pulsar J1024$-$0719, corresponding to the same typical noise realizations shown in Fig.~\ref{fig:toa_1024_both}. 
    \emph{Left panel:} Mostly internal noise scenario. Although the internal torque dominates the fluctuations, a small amount of external torque noise and the finite relaxation time $\tau$ lead to deviations from the red line of slope $\alpha = 1$ (the black line marks $\alpha = -x_p/x_n = -4$ for comparison).
    \emph{Right panel:} Mostly external noise scenario. Despite the dominance of external fluctuations, the residual internal noise and finite $\tau$ still cause slight deviations from the red slope line $\alpha = 1$. In both panels, the black line serves as a reference for $\alpha = -x_p/x_n = -4$.
    }
    \label{fig:corr_1024_both}
\end{figure*}

To examine whether the correlation slope $\alpha \approx \delta \Phi_n / \delta \Phi_p$ can be extracted from phase-resolved data, we analyze the correlation trajectories in both frequency and phase space. Fig.~\ref{fig:corr_1024_both} shows the trajectories in the $(\delta \nu_p, \delta \nu_n)$ and $(\delta \Phi_p, \delta \Phi_n)$ planes for the same noise realizations used in Fig.~\ref{fig:toa_1024_both}. In the mostly internal noise scenario (left panel), the trajectories exhibit noticeable scatter about the red reference line of slope $\alpha = 1$, due to the relatively long relaxation time with respect to the observational baseline.

In the mostly external noise case (right panel), the trajectories are more tightly clustered around the red line ($\alpha = 1$), reflecting a nearly coherent response of both components to the common external forcing. However, slight deviations remain due to the residual presence of internal noise and a finite relaxation timescale $\tau$. 
As in the canonical case, the black reference line with slope $\alpha = -x_p/x_n = -4$ is included for comparison, though it plays a minor role in the external noise regime.

Overall, the behavior observed in Figs.~\ref{fig:toa_1024_both} and \ref{fig:corr_1024_both} suggests that even in MSPs with high coupling and low glitch activity, measurable TN can arise from small stochastic fluctuations in the external torque. Moreover, while correlation slopes can, in principle, be recovered from residual trajectories, the accuracy of such measurements will depend critically on two ratios: how the relaxation timescale compares to the observation baseline and the amplitude ratio of internal and external torque fluctuations.

\section{GW detection prospects}
\label{gwdetection}

To determine whether joint electromagnetic (EM) and gravitational wave (GW) observations can be used to study the nature of the torques acting on a neutron star (NS) and the coupling between its components, it is crucial to assess whether the phase fluctuations discussed above are detectable, at least with the next generation of GW detectors.

To do this, we will consider the sensitivity of the planned next-generation ground-based GW detector, the Einstein Telescope (ET), considering in particular the configuration with two parallel 15 km long L-shaped detectors, as described in \citet{Branchesi_2023}. 
The detectability of the signal depends on the exact GW emission mechanisms, the distance, and the frequency of the source. However, it has been shown \citep{ETbluebook} that for the currently observed galactic neutron stars, several hundred would be observable after one year by ET, assuming emission due to a quadrupolar mountain, with a characteristic amplitude $h_0$ of the form
\begin{equation}
\label{mountains}
    h_0 = 10^{-25}\left(\frac{10\mbox{ kpc}}{d}\right)\left(\frac{\epsilon}{10^{-6}}\right)\left(\frac{\nu_s}{500 \mbox{Hz}}\right) I_{45} 
    \, ,
\end{equation}
where $d$ is the distance of the source, $\epsilon$ the ellipticity, $I_{45}$ the moment of inertia of the star in units of $10^{45}$ g cm$^2$, and $\nu_s$ its spin frequency (we recall that the GW frequency $\nu_{GW}=2\nu_s$ in this quadrupolar mountain scenario). 

If the source is detectable by ET, the error in measuring the phase $\delta\Phi_{GW}$ will be \citep{jones2004_timing}:
\begin{equation}
\delta\Phi_{GW} = \frac{N_{sim}}{\rho_d} \, ,
\label{mindelta}
\end{equation}
where $N_{sim}$ is a factor of order unity that depends on the number of parameters involved in the search. For our estimates, we use the conservative value of $N_{sim}=3$ (see \citealt{jones2004_timing} and references therein for a discussion of this point), and is $\rho_d$ the signal-to-noise ratio with which the signal is detected, calculated as~\citep{Watts08}
\begin{equation}\label{snr_coh}
\rho_d=h_0\sqrt{\frac{N_d T_{\rm obs}}{S_n(\nu)}} \, ,
\end{equation}
where $N_d$ is the number of detectors $T_{ \rm obs}$ is the observation length, and $S_n(\nu)$ the noise spectral density of the detector at frequency~$\nu$.

\begin{figure*} 
    \centering
    \includegraphics[width=0.6\textwidth]{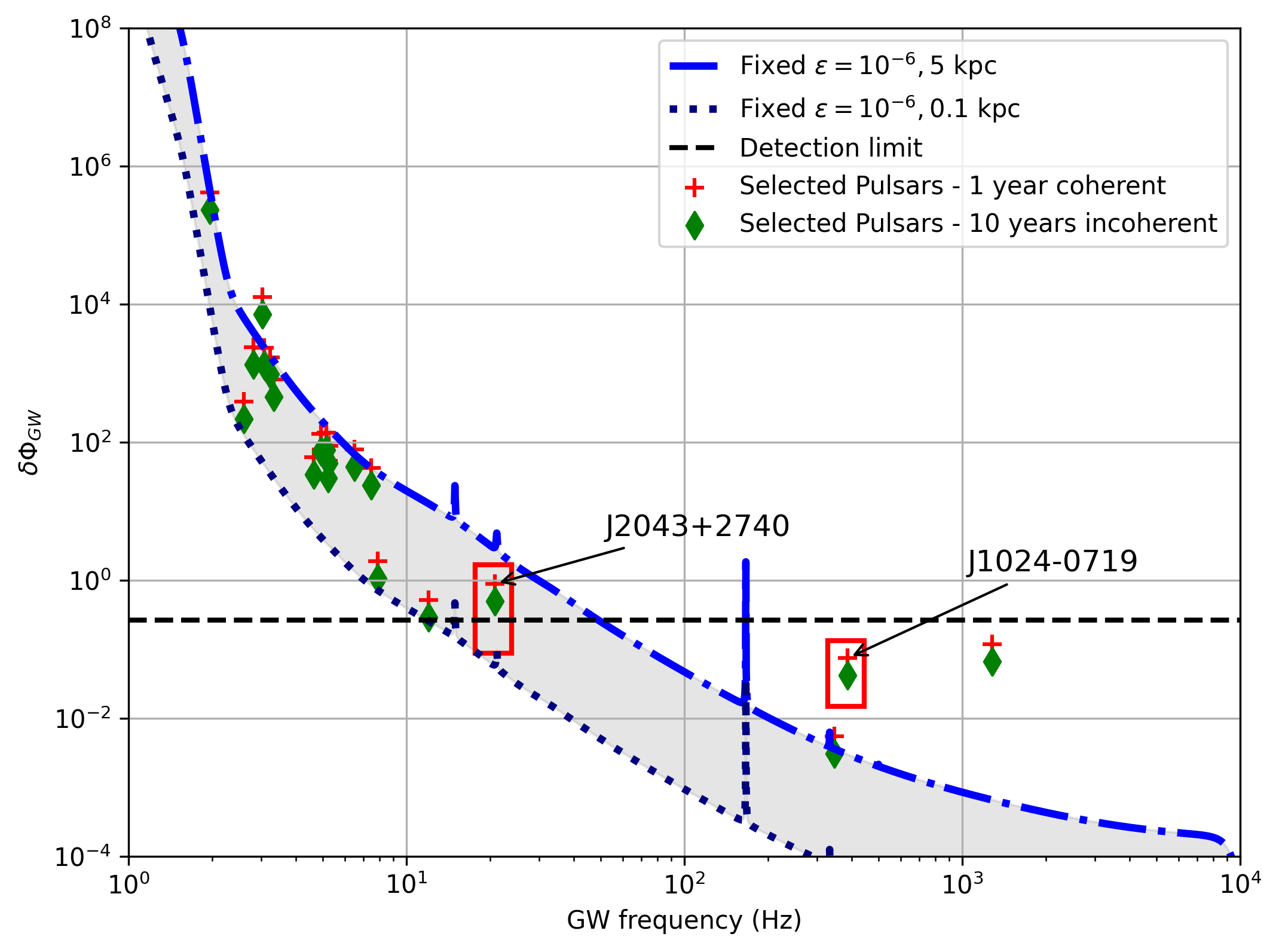}
    \caption{The minimum $\delta\Phi_{GW}$ detectable by ET with pulsar observations. The shaded region between the blue lines is for a 1-year coherent integration of a signal for sources with $\epsilon=10^{-6}$, with a distance between 0.1 kpc and 5 kpc. The crosses and diamonds are for the observed pulsars described in the text, considering both a 1-year coherent integration (crosses) and 10-year in-coherent integration (diamonds). The horizontal line represents the limit $\delta\Phi_{GW}=N/11.4 \approx 0.18$, which gives us an estimate of detectability. Pulsars above this line would not be detectable. Note, however, that pulsars below the line may very well have an intrinsic timing noise that is larger than the minimum detectable plotted here.}
    \label{coherent_1yr}
\end{figure*}

To measure a phase difference greater than the minimum in \eqref{mindelta} it is, of course, necessary for the source to be loud enough for detection, with louder sources allowing for more precise measurements of the TN. We set the limit for detection at $\rho_d > 11.4$~\citep{Watts08}, and consider two cases: first a 1 year coherent integration, and a $T_{\rm obs}=$ 10 yr total observation, summed incoherently over $T_{\rm coh}=$ 1  yr stretches of data, for which~\citep{Branchesi_2023}
\begin{equation}\label{snr_semicoh}
\rho_d\approx h_0\sqrt{\frac{N_d T_{\rm coh}}{S_n}} \,
\left( \frac{T_{ \rm obs}}{T_{\rm coh}} \right)^{1/4}  .
\end{equation}
In the following, we will consider the 2L configuration for ET \citep{Branchesi_2023}, and therefore take $N_d=2$, and one has $\rho_d^{10\text{yr}}\approx1.78\rho_d^{1\text{yr}} $.
With the above definitions, we can calculate the minimum phase wandering $\delta\Phi_{GW}$ that could be detected in a one-year observation, which is presented in Fig.~\ref{coherent_1yr} as a function of frequency. We plot our limits for a fixed reference value of $\epsilon=10^{-6}$ for all pulsars, for a range of distances between $d=0.1$ kpc and $d=5$ kpc, that would thus roughly encompass most observed pulsars. We see that for most galactic neutron stars, with a gravitational wave frequency around the KHz range, we would be able to measure, at best, phase wandering of the order of $\delta\Phi_{GW} \gtrsim 10^{-4}$rad.

We also show our limit for detection, which for our cutoff at signal-to-noise ratio $\rho_d=11.4$, gives us $\delta\Phi_{GW}<N/11.4\approx 0.18$ rad. This means that pulsars above this line would not be detectable with our assumptions on $\epsilon$. It does, however, indicate that for all detectable pulsars we would be able to measure a minimum $\delta\Phi<1\,$rad, therefore below the value that would cause our templates to drift out of phase with the real signal and hinder detection. We caution the reader, however, that we are simply estimating the smallest phase shift that could be measured. The shift in phase due to a specific pulsar's timing noise will depend on the strength of said noise, and as we have seen, it can be significantly larger.

In Fig.~\ref{coherent_1yr} we also show the minimum $\delta\Phi_{GW}$ that would be measurable by ET for a subset of observed pulsars, which are known to exhibit strong timing-noise \citep{Lyne+2013}: 17 non-recycled radio pulsars analysed by \cite{Shaw+2022, Lyne+2010}, and 3 millisecond pulsars that exhibit stronger timing-noise \citep{Perera+2019}, namely PSR J0437-4715, PSR J1024-0719 and PSR J1939+2134: ({ using eq. \ref{mindelta} and~\ref{snr_coh}})
\begin{equation}
    \delta\Phi_{GW} = \frac{2}{h_0 (d,\epsilon,\nu_s)} 
    \sqrt{\frac{S_n(2 \nu_s)}{2 \, T_{obs}}} \, ,
\end{equation}
where $I=10^{45}\,$g cm$^2$ is assumed for all objects.

Note that for the 17 standard radio pulsars, we assume a physically motivated ellipticity of $\epsilon = 10^{-6}$ for our analysis and use their measured distance. However, for the 3 millisecond pulsars, this value exceeds their spin-down upper limit, $\epsilon_{sd}$, which represents the ellipticity required for the observed spin-down to be entirely due to gravitational wave (GW) emission
\begin{equation}
    \epsilon_{sd}= 1.9\times 10^{-5} \!
    \left(\frac{100\,\mbox{Hz}}{\nu_s}\right)^{\frac{5}{2}} \!\!
    \left(\frac{\dot{\nu}_s}{10^{-10}\,\mbox{Hz/s}}\right)^{\frac{1}{2}} \!\!
    I_{45}^{-\frac{1}{2}} .
\end{equation}
For these 3 systems, we therefore assume GW emission occurs at their spin-down limit. Note that this smaller value of $\epsilon$ is the reason that the estimated minimum value of $\delta\Phi_{GW}$ that could be measured is larger than those in the shaded band.
We see that, despite the smaller ellipticities, the millisecond pulsars in our sample still represent the best targets for detection with ET, and would allow us to measure variations in phase due to timing noise of the order of $\delta\Phi_{GW}\approx 10^{-2}$rad. On the other hand, for our assumed mountain size of $\epsilon=10^{-6}$, only the fastest 'regular' pulsars would be loud enough for detection (and therefore allow to attempt a TN measurement). 

In fact, not only may the timing noise be detectable, but it may also be strong enough, in both of the pulsars we consider, to hinder detection, causing the signal to drift out of phase with our model by several tens of radians in the case of the millisecond pulsar PSR J1024-0719, for both internal and external noise models. This points towards the fact that timing noise must be accounted for in searches for these signals with ET, and mitigation strategies must be considered~\citep{Ashton2015}.

\section{Discussion}
\label{sezioncione_discussiolone}

Ongoing searches for CWs from isolated pulsars in LVK data, as well as future searches in ET data, require tackling both the technical problem of how to detect a signal in a realistic system for which there is spin wandering and what physics could be extracted from such a detection. The two quests are intertwined~\cite{jones2004_timing}, as hypotheses on the nature of the signal - in particular, its statistical properties that make it deviate from a purely deterministic one - allow, in principle, the optimization of the computationally expensive search procedures.  

Regarding the physics that may be extracted from future detection, we have assumed that pulsars also behave as `gravitational pulsars'. Hence, the natural idea would be to contrast the CW signal - which carries information on the global structure of the neutron star - with the EM one, which is likely to be influenced by the charged and non-superfluid components within the star.

Despite the exact physics of how a differentially rotating superfluid may impact the CW emission remaining an open issue, it is unlikely that the possibility of detecting a gravitational pulsar would strongly depend on this understanding. Rather, it could be that the observation of a gravitational pulsar will constrain such mechanisms and their interplay with the internal superfluid component.       
Therefore, we made the hypothesis that the CW signal follows a loose superfluid component within the star. 
This working assumption implies that CW measurements probe the body-averaged superfluid velocity based on the premise that the quadrupole moment is primarily influenced by the superfluid (this assumption can be relaxed, allowing for alternative scenarios to be considered within the same family of stochastic models discussed in~\citealt{meyers2021a}).
We then used a simple stochastic model for the CW signal and its EM counterpart to provide a first guideline for the interpretation of the data. 
In particular, within our null hypothesis, we highlighted how possible differences in the PSD of the EM and CW signals can be related to the presence of the loose superfluid component, its moment of inertia, and the global coupling timescale via the detection of corner frequencies. 
Similarly, the measurement of a certain degree of correlation $C$ between the CW signal and the EM counterpart can be linked to the nature of fluctuations: $C \approx 1$ would signal that timing noise fluctuations are rooted in the braking mechanism, whatever its nature, while $C<0$ is a signature of a fluctuating internal torque (likely to be related to classical or quantum turbulence or intermittent vortex creep, e.g.,~\citealt{MelatosLink2014,antonelli_timing2023}).

The level of timing precision achieved at radio wavelengths by current-generation radio telescopes is sufficient to significantly measure variations in timing residuals arising from timing noise. From our analysis, we find that millisecond pulsars (MSPs) are promising candidates for detecting CW frequency modulations induced by timing noise. Although MSPs are generally known as stable rotators, a small subset exhibits a noticeable amount of timing noise. Therefore, MSPs with substantial timing noise and rapid spin rates would be ideal sources for such detections.

Future-generation radio telescopes like SKA-Low and SKA-Mid will offer near-instantaneous sensitivity, enabling ultra-precise time-of-arrival measurements and the ability to detect a vast number of pulsars. The full Square Kilometre Array (SKA) has the potential to discover between 2,400 and 3,000 additional MSPs, depending on its exact sensitivity \citep{Keane_2014vja}. Based on the MSP selection criteria proposed by \citet{Halder_2023rfu} and the current pulsar population from the ATNF catalogue\footnote{
    \href{https://www.atnf.csiro.au/research/pulsar/psrcat/}{https://www.atnf.csiro.au/research/pulsar/psrcat/} 
} \citep{Manchester_2004bp}, we currently identify 572 MSPs. This suggests that, in the SKA era, the MSP population could increase by approximately fivefold. The sheer number of newly discovered millisecond pulsars, combined with their ultra-precise timing capabilities, will significantly enhance the pool of potential gravitational wave sources.

 

\paragraph{Funding Statement}
M.A. acknowledges support from the IN2P3 Master Project NewMAC,
the ANR project `Gravitational waves from hot neutron stars and properties of
ultra-dense matter' (GW-HNS, ANR-22-CE31-0001-01).
B.H. acknowledges support from the Polish National Science Centre grant OPUS 2023/49/B/ST9/02777 and OPUS LAP 2022/47/I/ST9/01494. 
A.B. acknowledges the support from the UK Science and Technology Facilities Council(STFC). 
A consolidated grant from STFC supports the pulsar research at the Jodrell Bank Centre for Astrophysics.









\appendix

\section{Fluctuations driven by Ornstein-Uhlenbeck noise}
\label{app1}

For completeness, we provide a more explicit description of the two-component model injected with Ornstein–Uhlenbeck (OU) noise. This allows us to provide a prototypical example for the claim in \eqref{psds} and to interpolate between the cases $\beta=0$ (when white noise is injected) and $\beta=2$ (when Wiener noise is injected).
To this end, we consider the 4-dimensional stochastic process
 \begin{equation}
 \label{OUinj}
 \begin{split}
     &\delta\dot{\mathbf{\Omega}}_t =  B \, \delta\mathbf{\Omega}_t + M V\mathbf{U}_t 
     =  B \, \delta\mathbf{\Omega}_t + M(\dot{\mathbf{W}}_t-\dot{\mathbf{U}}_t) \, ,
     \\
     &\dot{\mathbf{U}}_t =  -V \, \mathbf{U}_t +\dot{\mathbf{W}}_t \, , 
 \end{split}
 \end{equation}
where $M$ and $B$ are the ones in \eqref{2_comp_mat} and 
$\mathbf{W}_t=\left( W_\infty(t) , W_\mathcal{T}(t) \right)$ 
are two independent standard Wiener processes.  
Similarly, the two independent OU stochastic processes that drive the fluctuations in the internal and external noise are $\mathbf{U}_t= \left( U_\infty(t) , U_\mathcal{T}(t) \right)$. 
To ensure the independence of $U_\infty$ and $U_\mathcal{T}$, 
the matrix $V$ is diagonal,~$V=\text{diag}\left( \nu_\infty ,  \nu_\mathcal{T} \right)$.
The main difference with respect to the system in \eqref{langevin2} -- which is directly injected with white noise $\dot{\mathbf{W}}_t$ -- is that now we are injecting the noise $V \mathbf{U}_t$ that has -- in its stationary regime -- the PSD
\begin{equation}
    \text{PSD}\left( \nu_i U_i \right) = \dfrac{\nu_i^2}{\omega^2 + \nu_i^2}  \, ,
\end{equation}
where the frequencies $\nu_i$ are related to the physics of the process, leading to fluctuations that are temporally correlated over a timescale $1/\nu_i$.  
Therefore, injecting the noise model $V \mathbf{U}_t$ allows us to automatically retrieve the dynamics in \eqref{langevin2} in the double limit $\nu_i \rightarrow \infty$.
To check this statement more explicitly, we perform this double limit on the 
the full PSD $P_{p}(\omega)$ for the angular velocity residuals $\delta\Omega_p(t)$ of the normal component,
\begin{align}
 P_{p}(\omega)\propto &
    \frac{
    \tau^2  \sigma_\infty^2 \nu_\infty^2  } 
    {  \left(
    \nu_\infty^2  + \omega^2
    \right)
    \left( 1 + \omega^2 \tau^2
    \right)}+ \nonumber
    \\
    & + \frac{
    \tau^2  \sigma_\mathcal{T}^2 \nu_\mathcal{T}^2 } 
    { 
    \left(
    \nu_\mathcal{T}^2 + \omega^2
    \right)
    \left( 1 + \omega^2 \tau^2
    \right)}+ \nonumber
    \\
    & +\frac{ \sigma_\infty^2 \nu_\infty^2 x_p^2 }
    { \omega^2\left(
    \nu_\infty^2  + \omega^2
    \right)
    \left( 1 + \omega^2 \tau^2
    \right)} \, .
\label{eq:psdss}
\end{align}
\noindent
Explicit calculation of the double limit gives
\begin{equation}
    \lim_{\nu_i \rightarrow \infty} P_{p}(\omega) \propto
           \frac{
    \tau^2 \omega^2 \left( \sigma_\infty^2   + \sigma_\mathcal{T}^2   \right) 
+ \sigma_\infty^2   x_p^2  
} { \omega^2  \left( 1 + \omega^2 \tau^2 \right) } \, ,
\end{equation}
which is exactly the PSD for injected white noise in~\eqref{psds} for $\beta=2$.

Retrieving the limiting case of injected Wiener noise is more subtle. 
Roughly speaking, we should obtain this behaviour by considering the double limit $\nu_i\rightarrow 0$, which immediately implies $\dot{\mathbf{U}}=\dot{\mathbf{W}}$. However, the amplitude of noise in \eqref{OUinj} would also go to zero (i.e., the first equation would formally read $\delta\dot{\mathbf{\Omega}}_t =  B \, \delta\mathbf{\Omega}_t$). 
Therefore, we have to perform this double limit by keeping the ratio $\Sigma_i=\sigma_i/\nu_i$ constant.
Explicit calculation gives us
\begin{equation}
    \lim_{\nu_i \rightarrow \infty} P_{p}(\omega) \propto
           \frac{
    \tau^2 \omega^2 \left( \Sigma_\infty^2   + \Sigma_\mathcal{T}^2   \right) 
+ \Sigma_\infty^2   x_p^2  
} {  \omega^4  \left( 1 + \omega^2 \tau^2 \right) } \, ,
\end{equation}
that is consistent with~\eqref{psds} for $\beta=4$.

\bibliographystyle{pasa-mnras}
\bibliography{main}

\end{document}